\documentclass[a4paper,amsfonts, amssymb, amsmath,preprint, showkeys, nofootinbib, twoside]{revtex4-1}
\usepackage[english]{babel}
\usepackage{xcolor}
\usepackage{ulem}
\usepackage[utf8]{inputenc}
\usepackage[colorinlistoftodos, color=green!40, prependcaption]{todonotes}
\usepackage{amsthm}
\usepackage{mathtools}
\usepackage{amsmath}
\usepackage{physics}
\usepackage{xcolor}
\usepackage{graphicx}
\usepackage[left=23mm,right=13mm,top=35mm,columnsep=15pt]{geometry} 
\usepackage{adjustbox}
\usepackage{placeins}
\usepackage[T1]{fontenc}
\usepackage{lipsum}
\usepackage{csquotes}
\usepackage{amssymb}
\usepackage{amsfonts}
\usepackage{enumerate}
\usepackage{epstopdf}
\usepackage{makeidx}
\usepackage{caption}
\usepackage{subcaption}
\usepackage{fancyhdr}
\usepackage{marvosym}
\usepackage{wrapfig}
\usepackage{indentfirst} 
\def\up{\uparrow}
\def\down{\downarrow}
\def\bef{\begin{framed}}
\def\eef{\end{framed}}
\def\be{\begin{equation}}
\def\ee{\end{equation}}
\def\ber{\begin{eqnarray}}
\def\eer{\end{eqnarray}}

\def\sigmabold{\mbox{\boldmath $\sigma$}}

\def\ev{{\bf e}}

\def\Bv{{\bf B}}

\def\xv{{\bf x}}
\def\yv{{\bf y}}

\def\kv{{\bf k}}

\def\nn{\nonumber}

\usepackage[pdftex, pdftitle={Article}, pdfauthor={Author}]{hyperref} 
\begin{document}
\title{Photon Absorption of Two-dimensional Nonsymmorphic Dirac Semimetals}

\author{Amarnath Chakraborty}
\email{achakraborty@mail.missouri.edu}
\affiliation{Department of Physics and Astronomy, University of Missouri, Columbia, Missouri, USA}
\author{Guang Bian}
\affiliation{Department of Physics and Astronomy, University of Missouri, Columbia, Missouri, USA}
\author{Giovanni Vignale}
\email{vignaleg@missouri.edu}
\affiliation{Department of Physics and Astronomy, University of Missouri, Columbia, Missouri, USA}
\date{\today} 

\begin{abstract}
Two-dimensional Dirac semimetals have attracted a great deal of attention because of their linear energy dispersion and non-trivial Berry phase. These materials are rare because the nodal band structure is fragile against perturbations such as the spin-orbit coupling (SOC).  Recently, it has been reported that nonsymmorphic crystal lattices possess symmetry-enforced Dirac-like band dispersion around certain high symmetry momenta even in the presence of SOC. Here we calculate the optical absorption spectra of the nonsymmorphic semimetals, which hosts anisotropic Dirac cones, with different Fermi velocities along the $x$ and $y$ directions. Our calculations show that the optical absorption coefficient depends strongly on the anisotropy factor and the photon polarization. By rotating the latter, one can change the absorption coefficient by more than an order of magnitude, giving rise to {\it birefringence}. When a magnetic field is applied, the absorption coefficient also depends on an internal parameter, which we term the “mixing angle” of the band structure. This parameter becomes therefore accessible to experimental investigation.  We further find that an in-plane magnetic field, while leaving the system gapless, can induce a Van-Hove singularity in the joint density of states: this causes a significant enhancement of the optical absorption at the frequency of the singularity for one direction of polarization but not for the orthogonal one, making the optical properties even more strongly dependent on polarization and anisotropy. These results suggest that a very pure nonsymmorphic 2D Dirac semimetal can be an excellent candidate material for tunable magneto-optic devices.
\end{abstract}


\maketitle

\section{Introduction}
Over the past decade, two-dimensional (2D) Dirac-like electron gases have attracted tremendous research interest, with examples ranging from graphene \cite{Novoselov2005} to topological insulators \cite{RevModPhys.82.3045} to Dirac and Weyl semimetals.  These materials~\cite{3D,Yan,Nagaosa2020} possess several unique electronic and optical properties traceable to their linear energy dispersion and non-trivial Berry phase. Graphene has become the prototypical instance of two-dimensional Dirac fermions. However, the Dirac points in many existing 2D materials, including graphene, are vulnerable to spin-orbit coupling (SOC). Motivated by finding alternative 2D materials beyond graphene, various atomically thin materials, including silicene, germanene, few-layer black phosphorus, and other 2D compounds, have been theoretically proposed and experimentally prepared \cite{Sili,silicine,beyond,phosp}. Recently, symmetry-protected 2D Dirac semimetals have attracted intense interest. These materials feature Dirac points that are not gapped by SOC interaction and are protected by nonsymmorphic lattice symmetry \cite{2D,bismuth}. 

 Dirac-like band dispersions have recently been observed in the nonsymmorphic monolayer film, $\alpha$-bismuthene \cite{bismuth}. The lattice structure of $\alpha$-bismuthene belongs to the \#42 layer group ({\it pman}), as shown in Fig.~\ref{fig:image0}(a). There are two atomic sublayers marked by `A'  and `B' in $\alpha$-bismuthene with a vertical spacing of 3.02~\AA\ in between. The in-plane lattice constants are 4.53 and 4.72~\AA\ in the $x$ and $y$ directions, respectively. The lattice is invariant under a glide mirror reflection. That is a mirror reflection to the middle plane between the two sublayers followed by a translation by a half lattice constant in both $x$ and $y$ directions. This nonsymmorphic glide mirror symmetry leads to band degeneracy at the high symmetry momentum points $\bar{X}_{1}$ and $\bar{X}_{2}$ of the Brillouin zone, see Fig.~\ref{fig:image0}(b). The first-principles band structure (Fig.~\ref{fig:image0}(c)) indeed exhibits band crossing features among all the bands at  $\bar{X}_{1}$ and $\bar{X}_{2}$. The top Dirac nodes in valence bands $\bar{X}_{1}$ and $\bar{X}_{2}$ are denoted by `DP1' and `DP2', respectively. The two Dirac nodes are not connected by any lattice symmetry operation. Therefore, they are at different energies: DP1 at 0.7~eV and DP2 and 0.4~eV. Though the two nodes are not at the Fermi level in the pristine material, their energies can be shifted by applying a gating voltage and/or a strain. Another example of nonsymmorphic symmetry leading to protected Dirac points is found in Bi monolayer (\cite{bismuth}; see the section VII of the supplementary information \cite{SI} Fig S4), with  layer group  LG-p21/m11 (screw axis), where the Dirac points are predicted to be much closer to the Fermi level. 
 
 \begin{figure}
    \centering
    \includegraphics[width=1.0\linewidth]{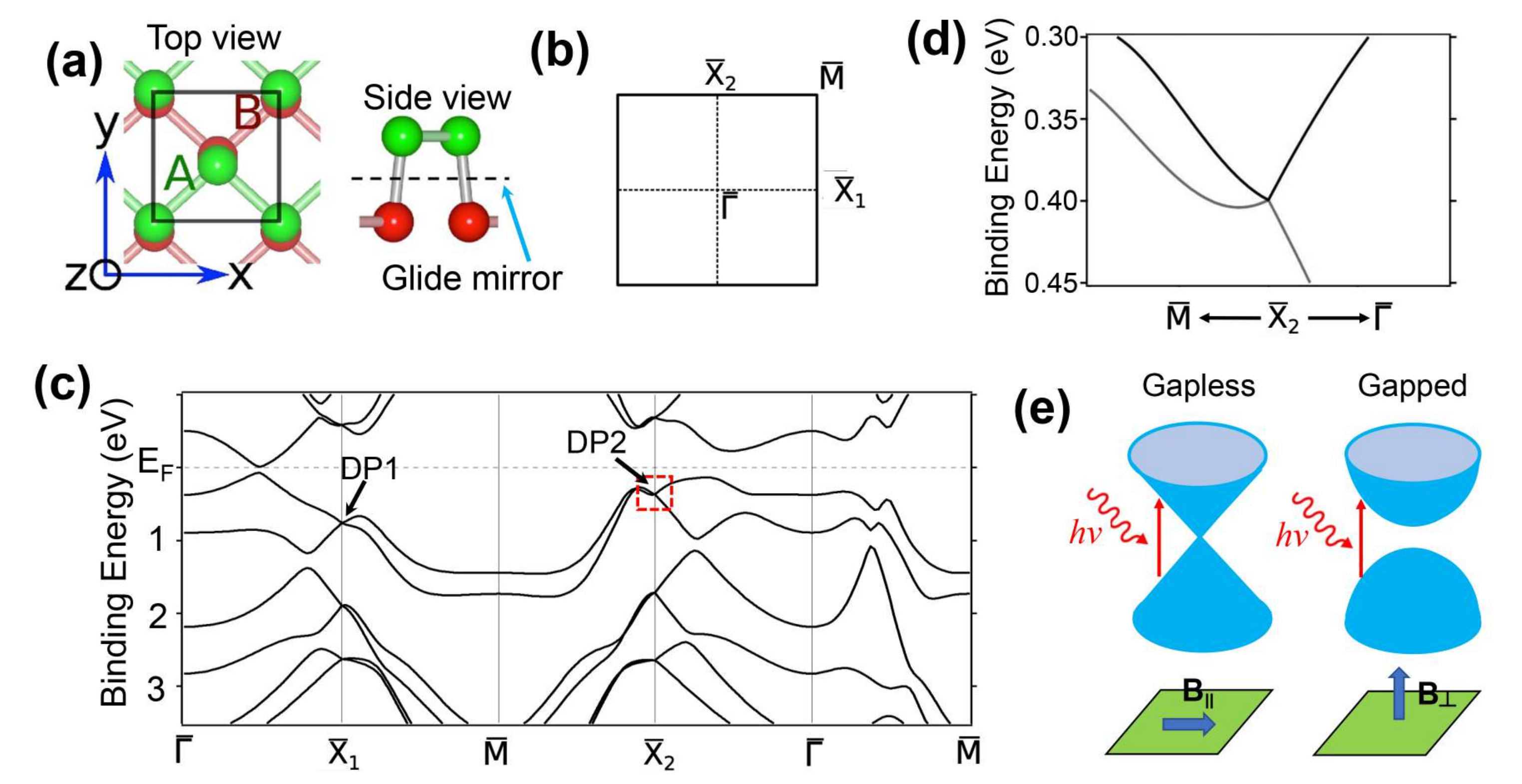}
\caption{\small \textbf{(a)} Lattice structure of $\alpha$-bismuthene. \textbf{(b)} Brillouin zone of $\alpha$-bismuthene where $\bar X_1$ ($\pi,0$) and $\bar X_2$ ($0,\pi$) are the two Dirac points (DP). \textbf{(c)} Band structure of $\alpha$-bismuthene. This shows both DPs are below Fermi energy. The Fermi velocities are $v_{x} = 3.95\times10^{5}\ m/s$ and $v_{y} = 2.12\times10^{5}\ m/s$ at DP1, and  $v_{x} = 1.19\times10^{5}\ m/s$ and $v_{y} = 4.67\times10^{5}\ m/s$ at DP2. \textbf{(d)} Zoom-in band structure marked by the red box in (c), emphaiszing the strong anisoptrpy present in DP2 ($\bar X_2$). \textbf{(e)} The band is gapped in case the direction of magnetic field is normal to the plane and gapless for parallel to the plane.}
\end{figure}\label{fig:image0}
 
 To effectively optimize and utilize the unique properties of 2D materials, various strategies have been proposed to tune the optical and electronic properties, such as the introduction of  electric fields, strain modulation \cite{tune,tune2}, atom doping \cite{Zhao2019}, strain engineering \cite{nanolett}, etc. However, to date, the optical properties of 2D nonsymmorphic Dirac materials have not been systematically investigated \cite{HfGeTe,Phosphorene,Universal}.
 
 There are two crucial differences between nonsymmorphic Dirac semimetals and graphene-like 2D Dirac materials. First, an anisotropy factor $\rho$ is allowed by the lower symmetry of the system. Our calculations show that $\rho$ plays a vital role in controlling photon absorption. Second, the spin (the $\sigma$ matrices) and orbital (the $\tau$  matrices) degrees of freedom are coupled together in our model Hamiltonian. Therefore, nonsymmorphic Dirac semimetals support strong spin-orbit coupling, unlike graphene, in which the spin and orbital parts are largely decoupled. These two facts yield a richer spectrum of optical properties in nonsymmorphic Dirac semimetals than in graphene, in particular a great sensitivity to the application of a magnetic field as discussed below.

In this paper, we will investigate the new features that nonsymmorphic symmetry brings to the optical absorption spectrum. Thus, we will assume that at least one nonsymmorphic Dirac node is present at the Fermi level and is directly accessible to photon absorption processes.  Our model for nonsymmorphic 2D Dirac cone is spelled out in Eq.~(\ref{Hamiltonian}) and we will examine the role played by the intrinsic parameters of that model, i.e., the "anisotropy factor" and the "mixing angle". While the Dirac cone is expected, under this assumption, to be the main contributor to the low-frequency optical absorption spectrum, it is not possible, in general,  to completely eliminate low-frequency contributions from metallic regions of momentum space, where the Fermi level crosses partially occupied bands.  This point will be addressed in detail in the concluding section, where we will argue that the residual metallic absorption can be clearly separated from  Dirac-cone absorption in  sufficiently clean samples, because it gives rise to a distinct Drude absorption peak.

We perform our calculations without and with the magnetic field, and in the latter case, we consider three orthogonal directions of the field. We find that a magnetic field perpendicular to the plane opens a gap in the spectrum. The magnitude of this gap can be related to the internal "mixing angle" -- a quantity not directly accessible from the band structure in the absence of a magnetic field. On the other hand, an in-plane magnetic field leaves the system gapless but splits the Dirac nodes. For one direction of the magnetic field, a Van-Hove singularity appears in the joint density of states. It is associated with a change in the topology of the constant energy contours at a saddle point in the band structure.  The logarithmic divergence of the joint density of states leads to enhanced absorption for one direction of the photon polarization but not for the orthogonal one, which implies that the absorption coefficient is very sensitive to the polarization of the incident light when the frequency approaches the Van-Hove singularity.

One of our significant findings is that the absorption coefficient can be tuned by changing the polarization and frequency of the incoming wave. This tunability is further enhanced by the intrinsic anisotropy of the Dirac cones.  These results open the door to interesting magneto-optical applications of 2D nonsymmorphic Dirac materials.

\section{Model and symmetries}
 We have taken the example of $\alpha$-bismuthene as the nonsymmorphic material described in the reference \cite{bismuth}. According to this paper the Dirac cones exists at $\bar X_1 = (\pi,0)$ and $\bar X_2=(0,\pi)$ of the Brillouin zone. $\alpha$-Bi is
nonmagnetic and centrosymmetric, so the time-reversal (T)
and inversion (P) symmetries are preserved. The symmetries in this model can be described by the three generators:
    \begin{equation}\label{generators}
        \begin{aligned}
        \tilde M_z: (x+1/2,y+1/2,-z)\ i\sigma_z\\  
        P: (-x,-y,-z)\ \sigma_0\\
        M_x:(-x,y,z)\ i \sigma_x
    \end{aligned}
    \end{equation}
    
Here, $\tilde M_z$ represents a nonsymmorphic glide mirror operation -- the mirror reflection accompanied by a half lattice translation parallel to the mirror plane in the case of $\alpha$-bismuthene. In Eq.~(\ref{Hamiltonian}), x, y z are spatial coordinates while $\sigma_i$ are Pauli matrices for the spin degree of freedom. Equation (\ref{generators}) describes the action of symmetry operators on the spatial coordinates and spin space. It would instead represent a screw-axis symmetry in the case of monolayer Bi. We can write the  matrix representation of the symmetry operators as 
\begin{equation}
    \begin{aligned}
        T &=-i\sigma_y\otimes\tau_0 K \ \text{(Time reversal)}\\ 
 \tilde M_z&= \sigma_z\otimes\tau_y \ \text{(Glide mirror)}\\
        P &= \sigma_0\otimes\tau_x \ \text{(Parity)}\\
      M_x &=-i \sigma_x\otimes\tau_x\ \text{(Mirror -x)},
    \end{aligned}
\end{equation}
where $K$ is the complex conjugation part which can be thought of a $2 \times 2$ matrix. Here, $\sigma_i$ $(i=x,y,z)$ are Pauli matrices for the spin and $\tau_i$ $(i=x,y,z)$ are the Pauli matrices for orbital degrees of freedom respectively and $\sigma_0$ and $\tau_0$ are the 2$\times$2 identity matrix \cite{bismuth}.     

\subsection{Model Hamiltonian without magnetic field}
For our purpose, we first consider the Dirac cone $\bar X_1 = (\pi,0)$ where the Hamiltonian can be written as:
\begin{equation}\label{Hamiltonian}
    \begin{split}
      H &= \rho v k_x(\cos\alpha \ \sigma_x\otimes\tau_z +\sin\alpha \ \sigma_0\otimes\tau_y)+v k_y\ (\sigma_y\otimes\tau_z)\\
      & =\rho vk_x(\cos\alpha \ \gamma_z +\sin\alpha \ \gamma_x)+v k_y \gamma_y \ ,
    \end{split}
        \end{equation}
where $\gamma_z=\sigma_x\otimes\tau_z$, $\gamma_y=\sigma_y\otimes\tau_z$ and $\gamma_x=\sigma_0\otimes\tau_y$. We define $\rho$ as the anisotropy factor $\rho = \frac{v_x}{v_y}$ because it refers to the mismatch in the Fermi velocity along $x$ and $y$-direction, where $v_y=v$ and $\rho v = v_x$. The angle $\alpha$ is what we termed the "mixing angle," which is an intrinsic parameter of the model as well. \\
The $\tilde M_z$ symmetry helps us to decompose the Hamiltonian into two 2$\times$2 matrices representing the $\tilde M_z$ even sector (eigenvalue +1) and odd sector   (eigenvalue -1). For convenience we chose the even sector to work with, corresponding to $\sigma_z =1$, $\tau_y=1$ and $\sigma_z=-1$, $\tau_y=-1$. Choosing a convenient basis (Please check the section I \cite{SI}) in the even sector we arrive at the reduced form of the Hamiltonian:
\begin{equation}\label{reduced_ham}
    \tilde H_{\tilde M_z=1}= \rho v k_x (\sigma_x \cos \alpha  \ +  \sigma_y \sin \alpha) + v k_y\sigma_z,
\end{equation}
The anisotropy factor $\rho$ is material-dependent. The two Dirac cones of bismuthene have different $\rho$ values, namely, $\rho$ = 1.86 at $\bar X_1$ and 0.25 at $\bar X_2$, because the two valleys are not connected by any crystal symmetries. In the following discussion, we will take $\rho$ as a free parameter of the model and study the $\rho$-dependence of optical absorption.

In both cases, the eigenvalues are respectively.
\begin{equation}
    E=\pm v\sqrt{\rho^2k_x^2+k_y^2}
\end{equation}

Notice that we could reduce the whole Hamiltonian to 2$\times$2 form only because the system has the glide mirror symmetry $\tilde M_z$. 
\subsection{Model Hamiltonian with Magnetic field}
Introducing a magnetic field in this system reduces the symmetry. We go back to the original form of the Hamiltonian to find the eigenstates and eigenvectors.
The Hamiltonian with the magnetic field is
\begin{equation}
      \tilde H= \rho v k_x \big(\cos \alpha \ \gamma_z +\sin \alpha \ \gamma_x \big)+v k_y \gamma_y + \Vec{B} \cdot \tau_{0} \ \Vec{\sigma},
  \end{equation}
The direction of the magnetic field is described by polar ($\mu$) and azimuthal angles ($\nu$) such that
\be
\Vec{B}= B \left( \sin \mu  \cos \nu , \sin \mu \sin \nu, \cos \mu \right),
\ee
We also set
\be
v\rho k_x = \bar k \cos \phi\,,~~~~~ vk_y =  \bar k \sin \phi\,.
\ee
With this notation the energy eigenvalues are 
\footnote{Notice that the eigenvalues come in pairs of opposite sign. This is because the original Hamiltonian anticommutes with the operator $\tau_z \sigmabold\cdot(\tilde \kv\times \Bv)$ where $\tilde \kv$ is a vector with components  $(k_x\sin\alpha,k_y,0)$.}
\be\label{Ephi}
\bar E(\phi)=\pm \sqrt{B^2+\bar k^2 \pm2 B \bar k F(\phi)}\,,
\ee
where 
\be
F(\phi)=\sqrt{(\cos \phi\cos \alpha)^2+\sin^2\mu (\cos \phi \cos \nu \sin \alpha+\sin \phi \sin \nu)^2}\ ,
\ee
The spectrum is generally gapped in presence of magnetic field. We find the band gap $\Delta$ using Eq.~(\ref{Ephi}) as:
\be\label{Gap1}
\frac{\Delta}{B} = 2\min \frac{|\bar E(\phi)|}{B}=2\sqrt{1-\frac{A+C+\sqrt{A^2+C^2+2AC\cos(2\gamma)}}{2}} ,
\ee
where we have defined
\be\label{Gap2}
\cos (2\gamma)=\frac{\sin^2\alpha-\tan^2\nu}{\sin^2\alpha+\tan^2\nu}\,,
\ee
and
\be\label{Gap3}
A=\cos^2\alpha\,,~~~~C= \sin^2\mu(\cos^2\nu\sin^2\alpha+\sin^2\nu).
\ee

In the special case of magnetic field along the $z$-axis ($\mu=0$), we get $A=\cos^2\alpha$, $C=0$, $\frac{\Delta}{B}=2\sin\alpha$.

If the magnetic field is along the $x$-axis ($\mu=\pi/2,\nu=0$), we have  $A=\cos^2\alpha$, $C=\sin^2\alpha$, $\cos (2\gamma)=1$, yielding $\Delta=0$.

If the magnetic field is along the $y$-axis ($\mu=\pi/2,\nu=\pi/2$), we have  $A=\cos^2\alpha$, $C=1$, $\cos (2\gamma)=-1$, yielding $\Delta=0$.

The bandgap along $z$-direction of field depends on the mixing angle $\alpha$. Thus, by measuring the bandgap in the presence of a magnetic field one can determine the value of the mixing angle.  
\section{Mathematical framework to calculate the optical absorption coefficient} \label{sec:math}
The electromagnetic interaction is given by $\Vec{J} \cdot \Vec{A}(t)$,
where $\vec J =\frac{\partial H}{\partial \Vec{k}}$ is the current operator with components
\be
J_x =\rho v (\cos \alpha \ \sigma_x\otimes\tau_z +\sin\alpha \ \sigma_0\otimes\tau_y)\,,~~~J_y=v\tau_z\sigma_y\,.
\ee
and $\vec A(t)$ is the vector potential 
\be
\vec A(t)=A_0  \cos (\omega t)\ \ev ,
\ee
where $A_0$ is the amplitude of the electromagnetic wave and the polarization vector $\ev$ is defined as.~\footnote{We are working in low energy regime thus using the dipole approximation and neglecting the photon momentum.}
\be\label{genpol}
\ev =\cos\beta \hat\xv+e^{i\delta}\sin\beta \hat \yv
\ee
The optical transition probability is 
\begin{equation}\label{genmat}
        W_{v\to c}(\omega)=
        \frac{A^2v^2}{2\pi\hbar}\sum_j \int_0^\infty d\bar k\bar k \int_{\phi_{\rm min}}^{\phi_{\rm max}} d\phi \ |M_{j}(\phi)|^2\frac{\delta(\bar k-\bar k_{j}(\phi))}{|d(E_{c}-E_{v})(\bar k,\phi)/d\bar k|},
\end{equation}
where  $\bar k_j(\phi)$, with $j=+$ or $-$ are the constant energy contours at the transition energy $\omega$, i.e., the solutions of the equation $E_{c}(\bar k,\phi)-E_{v}(\bar k,\phi)=2 \hbar \omega$, and $v \to c$ refers to the transition from the valence band to the conduction band. The squared matrix element of the current operator is given by
\begin{equation}
    \begin{aligned}
        |M_{j}(\phi)|^2 \equiv\left\vert\langle\psi_{c j}(\phi)|\rho^{1/2}\  \cos \beta \ J_x+e^{-i\delta}\rho^{-1/2}\ \sin \beta \ J_y|\psi_{v j}(\phi)\rangle\right\vert^2\,,
        \end{aligned}
\end{equation}
where
\be
 |\psi_{v j}(\phi)\rangle \equiv |\psi_{v}(\bar k_{j}(\phi),\phi)\rangle\,,~~~|\psi_{c j}(\phi)\rangle \equiv |\psi_{c}(\bar k_{j}(\phi),\phi)\rangle\\ .
\ee
are, respectively, the valence and conduction band states evaluated at the iso-energy surface, i.e., $\bar k=\bar k_j(\phi)$. From Eq.~(\ref{genmat}) we have an expression for the transition probability that we can utilize to calculate the optical absorption from different pairs of bands (e.g, $v_{1}\to c_{1}$ or $v_{1}\to c_{2}$), with or without the magnetic field. 

For example, if we consider only intra-valley transition $v_{1}\to c_{1}$ we find that the transition probability is
\be
W_{v1\to c1}(\omega)=\frac{A^2\omega}{2\pi\hbar}\int_{\phi_{\rm min}}^{\phi_{\rm max}}d\phi \frac{k_+(\phi)|M_{+}(\phi)|^2+k_-(\phi)|M_{-}(\phi)|^2}{\sqrt{F(\phi)^2 + \left (\frac{\omega}{2}\right)^2-1}}\ ,
\ee
where $\omega$ is the excitation energy and has value $\Delta<\omega<2B$ \ ($\Delta$ is the bandgap of the system). Here $\phi_{\rm min}$ and $\phi_{\rm max}$ are the lower and upper limits of integration, defined by the solutions of the equation
\be\label{fphi}
F(\phi)^2=1-\left (\frac{\omega}{2}\right)^2\,.
\ee  
The two branches of the energy contours $\bar k_\pm(\phi)$ coincide at $\phi_{\rm min}$ and $\phi_{\rm max}$ and thus combine to produce closed contours around the minimum excitation energy (Look into the supplementary material section III  \cite{SI} for calculation of the iso-energy contours).

\section{Results and Discussions} 
Based on the mathematical framework described before, we can calculate the optical absorption coefficient for a general direction polarization with and without the magnetic field. First, we start with the case where there is no magnetic field in our system, and later we introduce a magnetic field in different directions. 

\subsection{Optical absorption coefficient without the magnetic field}
In this case, as there is no magnetic field present, we have used the reduced Hamiltonian Eq.~(\ref{reduced_ham}) for $\tilde M_z=+1$. The glide mirror symmetry holds for this case and has distinct eigenvalues for the two valleys ($\tilde M_z=\pm 1$), which prevents  inter-valley transitions. As a result, we end up with only intra-valley transitions (i.e., $v_{1}\to c_{1}$ or $v_{2}\to c_{2}$ which are exactly equal). For the valley $\bar X_{1}$, we get the transition probability
\be\label{nofield}
  W_{v1\to c1}=\frac{A^2\omega}{8}\Big(\rho \ \cos^2 \beta + \rho^{-1} \sin^2 \beta \Big),
\ee
The optical absorption coefficient calculated \cite{katsnelson_2012} for general direction of polarization  without magnetic field takes the form:
\begin{equation}
    \eta= \frac{\pi  e^2}{2 \hbar c} f(\rho, \beta)\,,
\end{equation}
where we have defined  the {\it birefringence} function 
\begin{equation}\label{biref}
    f(\rho, \beta)= \left(\rho \ \cos^2{\beta}\ +\ \rho^{-1}\ \sin^2{\beta}\right)\,.
\end{equation}
We have reinstated the $\hbar$ in order to get the dimensionless quantity, the fine structure constant ($\frac{e^2}{\hbar c}=1/137$). 
This expression of optical absorbance is calculated for a single  valley (which in the case of $\alpha$-bismuthene  could be either the $\bar X_1$ or the $\bar X_2$ valley) as a function of the anisotropy factor $\rho$. It is also evident that this result is very similar to the well-known result for graphene, 
\footnote{The absorbance for monolayer graphene is $\eta_{graphene} = \frac{\pi e^2}{\hbar c}$.} 
in that the absorption coefficient is frequency-independent, but now it has a strong dependence on polarization, as indicated by the {\it birefringence} function. For large anisotropy ($\rho\ll1$ or $\rho\gg 1$) the absorption coefficient can change by a significant factor upon rotating the polarization angle.  

\subsection{Optical absorption coefficient in the presence of a magnetic field along the z-direction}
When working with the magnetic field along the $z$-direction, the Zeeman term preserves the glide mirror symmetry; thus, we can still decouple the Zeeman term for even and odd sectors under the glide mirror symmetry operator. The Zeeman term $B \tau_0 \sigma_z$ can be reduced to $B \sigma_z$ in the even sector and $-B \sigma_z$ in the odd sector. We have used the reduced Hamiltonian for the even sector of $\tilde{M_{z}}$ (Eq.~(\ref{reduced_ham})) and added the Zeeman term:
 \begin{equation}\label{Ham_z}
   H'_{M_z=1}=(\rho v k_x \cos \alpha+B )\sigma_x\ + \ \rho v k_x \sin \alpha \sigma_y\ +\ v k_y \sigma_z \ ,
\end{equation}
We write the eigenvalues as
    \begin{equation}\label{eig_z}
  E=\pm v\sqrt{\left( \rho k_x + \frac{B}{v} \cos\alpha \right)^2+k^2_y+\frac{B^2}{v^2} \sin^2\alpha}\ .
\end{equation}

\begin{figure}
    \centering
    \includegraphics[width=1.0\textwidth]{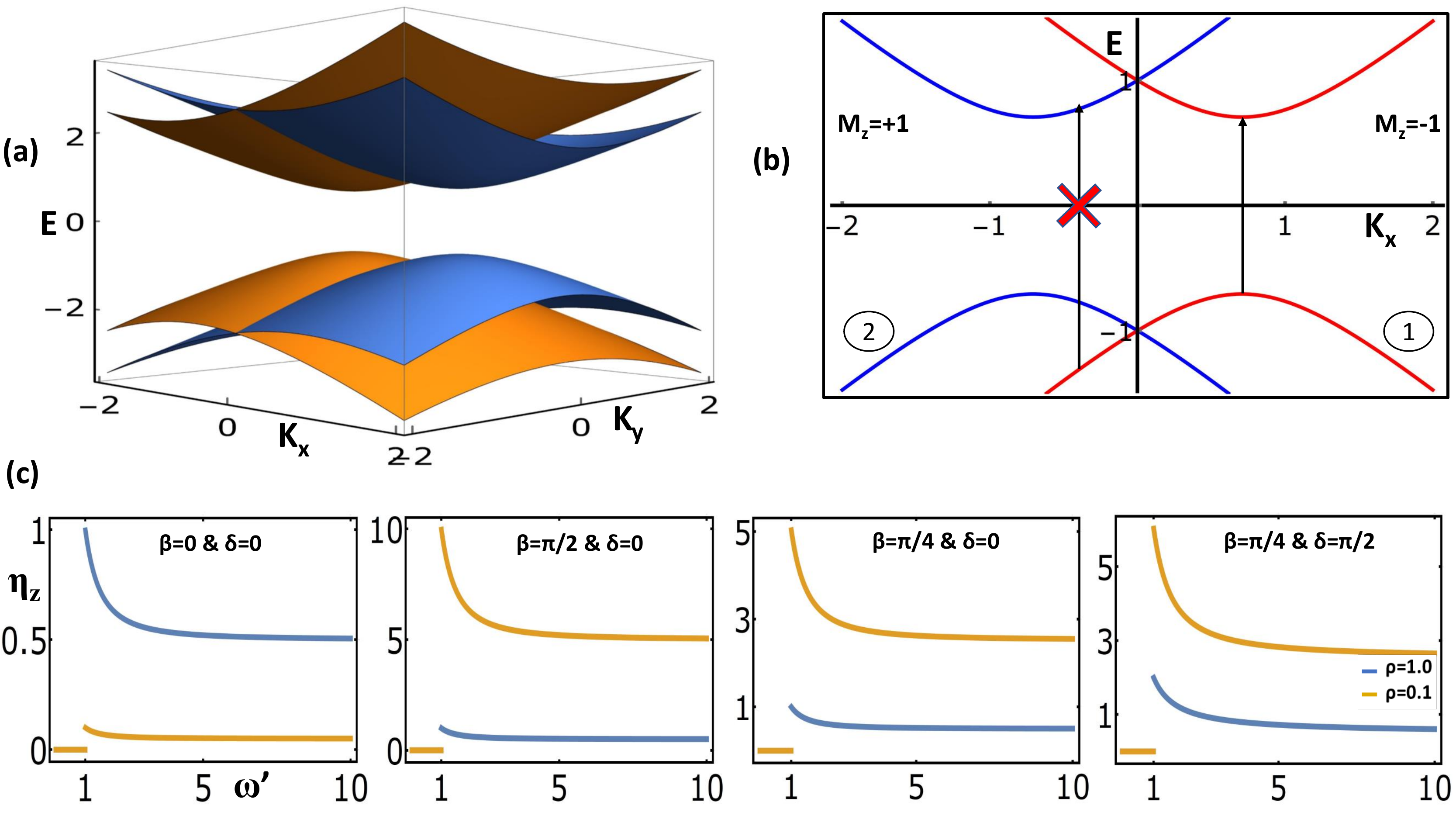}
\caption{\small \textbf{(a)} Three-dimensional plot of the band dispersion for magnetic field in the $z$-direction. Notice that there is a band gap $\Delta=2B \sin\alpha$ at two points separated by a wave vector proprtional to $B_z$ along $k_x$. The momenta $k_x$ and $k_y$ are in units of $\frac{B}{v}$ and the energy is in units of B. \textbf{(b)} This figure shows that only intra-valley transitions are possible as the two valleys hold opposite eigenvalues for $\tilde M_z$. \textbf{(c)}  Plot of the optical absorption coefficient vs scaled frequency $\omega'=\frac{\omega}{2B \sin\alpha}$ for four different polarizations. The plot starts at $\omega\geq 2B \sin\alpha$.}
\label{fig:image1}
\end{figure}

From Fig.~\ref{fig:image1}(a), we see the energy bands when the magnetic field is along the $z$-direction, and there exists a gap equals to $\Delta=2B \sin\alpha$ (Eq.~(\ref{Gap1})), which means the bandgap has dependence over the intrinsic quantity "mixing angle".

Both the current operators $J_x$ and $J_y$ and our Hamiltonian (Eq.~(\ref{Ham_z})) commutes with $\tilde M_z$, which makes the transitions preserve the parity of $\tilde M_z$. In Fig.~\ref{fig:image1}(b), we have shown the two valleys have opposite symmetries against the glide mirror symmetry for the $z$-direction of the magnetic field, and hence inter-valley transitions are not allowed.

Calculations similar to the previous section for optical absorbance ($\eta_z$) for the $\bar X_1$ valley leads us to
\begin{equation}\label{etazz}
    \eta_z=\frac{\pi e^2}{2\hbar  c} \Bigg[f(\rho,\beta) (1+\frac{4B^2}{\omega^2}\sin^2\alpha)+\frac{4B}{\omega}  \sin\alpha  \sin\delta  \sin2\beta\Bigg]\ .
\end{equation}
The {\it birefringence} function that appears in this expression will let us control the absorbance with the polarization of light. Also, it has a dependence on the frequency. For the $\bar X_2$ valley the expression is exactly the same with a different value for the anisotropy factor.
To understand the effect of the anisotropy factor on the absorbance, we assumed one Dirac point to be nearly isotropic ($\rho \approx 1.0$) and the other Dirac point to be highly anisotropic ($\rho \approx 0.1$). In Fig.~{\ref{fig:image1}}(c), we have plotted the optical absorption coefficient (Eq.~\ref{etazz}) with frequency (scaled frequency $\omega'=\frac{\omega}{2B \sin\alpha}$) for both the isotropic and anisotropic Dirac points for four different polarization. From these figures, one can see when the polarization is along $x$-direction ($\beta=0;\delta=0$) (Fig.~{\ref{fig:image1}}(c(i)) the dominating Dirac point is the isotropic one ($\rho = 1.0$) unlike the other cases where the the anisotropic ($\rho = 0.1$) case prevails in the absorption spectrum.

It shows when we change the polarization to the $x$-direction ($\beta=0;\delta=0$), we can significantly decrease the total absorbance, where if we switch to $y$-polarisation ($\beta=\pi/2;\delta=0$) we increase the total absorption by an order of magnitude. This shows the Dirac point mainly contributing to the enhancement of the absorbance is the anisotropic one, and this is due to the {\it birefringence} function present in the expression.

\subsection{Optical Absorption Coefficient in the presence of the Magnetic Field along the x-direction}
\begin{figure}
    \centering
    \includegraphics[width=1.0\linewidth]{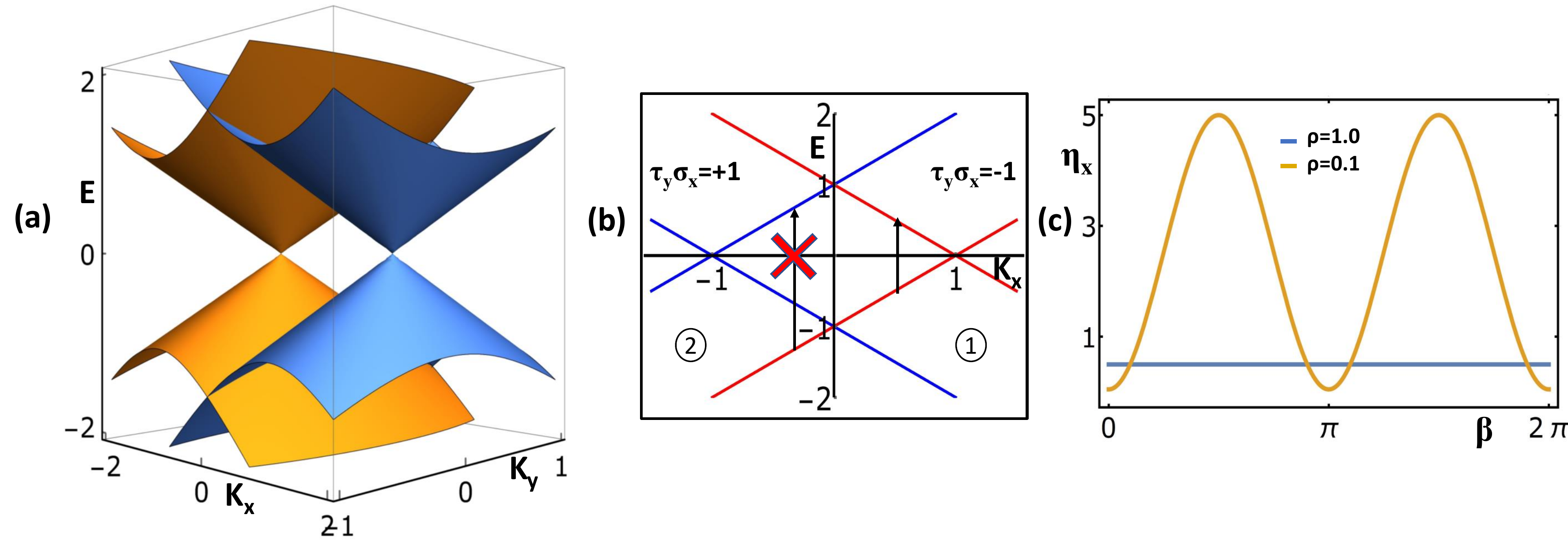}
    \caption{\small \textbf{(a)} Three-dimensional plot of the energy bands for magnetic field along the $x$-direction. The spectrum is gapless. \textbf{(b)} This figure shows that only intra-valley transitions are possible as the two valleys hold opposite eigenvalues of $\tau_y \sigma_x$. \textbf{(c)} Plot of the absorption coefficient $\eta_x$ vs $\beta$ for $\alpha = \pi/4$, for two different values of $\rho$, referring to the isotropic ($\rho=1.0$) and the anisotrpic ($\rho=0.1$) valley respectively. The absorption coefficient is in a unit of $\eta_{grahene}$.}
\label{fig:image2}
\end{figure}

For the magnetic field at $x$-direction, the Hamiltonian can be written as
\begin{equation}
  H= \rho v k_x \ \cos \alpha \ \gamma_z + \rho v k_x \ \sin \alpha\ 
  \gamma_x + v k_y \gamma_y + B_x\ \tau_{0}\sigma_x \ ,
\end{equation}
The corresponding eigenvalues are:
\begin{equation}\label{eig_x}
    E=\pm \sqrt{(\rho v k_x \pm B)^2 + v^2 k_y^2} \ ,
\end{equation}
From Fig.~\ref{fig:image2}(a), we see that there is no bandgap, and the two valleys cross each other. Similarly to the $z$-direction, we find that a symmetry operator $\tau_y \sigma_x$ is present in this case, which commutes with the current operators and the Hamiltonian. This has opposite signs in the two valleys (shown Fig.~\ref{fig:image2}(b)) and thus only allows intra-valley transitions for optical absorption. This symmetry operator  $\tau_y \sigma_x$ (product of the two broken symmetries, time reversal, and glide mirror) protects Dirac cones even though time-reversal symmetry is broken due to the presence of the magnetic field.  

For each valley, we get the optical absorption coefficient in the form:
\begin{equation}\label{eq:eq18}
    \eta_x= \frac{\pi e^2}{2\hbar c}\   f(\rho,\beta)\,,
\end{equation}
which coincides with the result obtained with no magnetic field. 

\subsection{Optical absorption coefficient in the presence of a magnetic field in the y-direction}

For a magnetic field in the $y$-direction, the Hamiltonian can be written as
\begin{equation}
  H= \rho v k_x \cos \alpha  \gamma_z + \rho v k_x  \sin \alpha
  \gamma_x + v k_y \gamma_y + B_y \tau_{0} \sigma_y\,,
\end{equation}
where the $\gamma$ matrices are defined after Eq.~(\ref {Hamiltonian}). Its eigenvalues are
\begin{equation}\label{eig_y}
    E=\pm \sqrt{\big(\sqrt{(\rho v k_x \cos \alpha)^2+ (vk_y)^2} \pm B\big)^2 + \big(\rho v k_x \sin \alpha \big)^2}\,.
\end{equation}

\begin{figure}[h!]
    \includegraphics[width=1.0\linewidth]{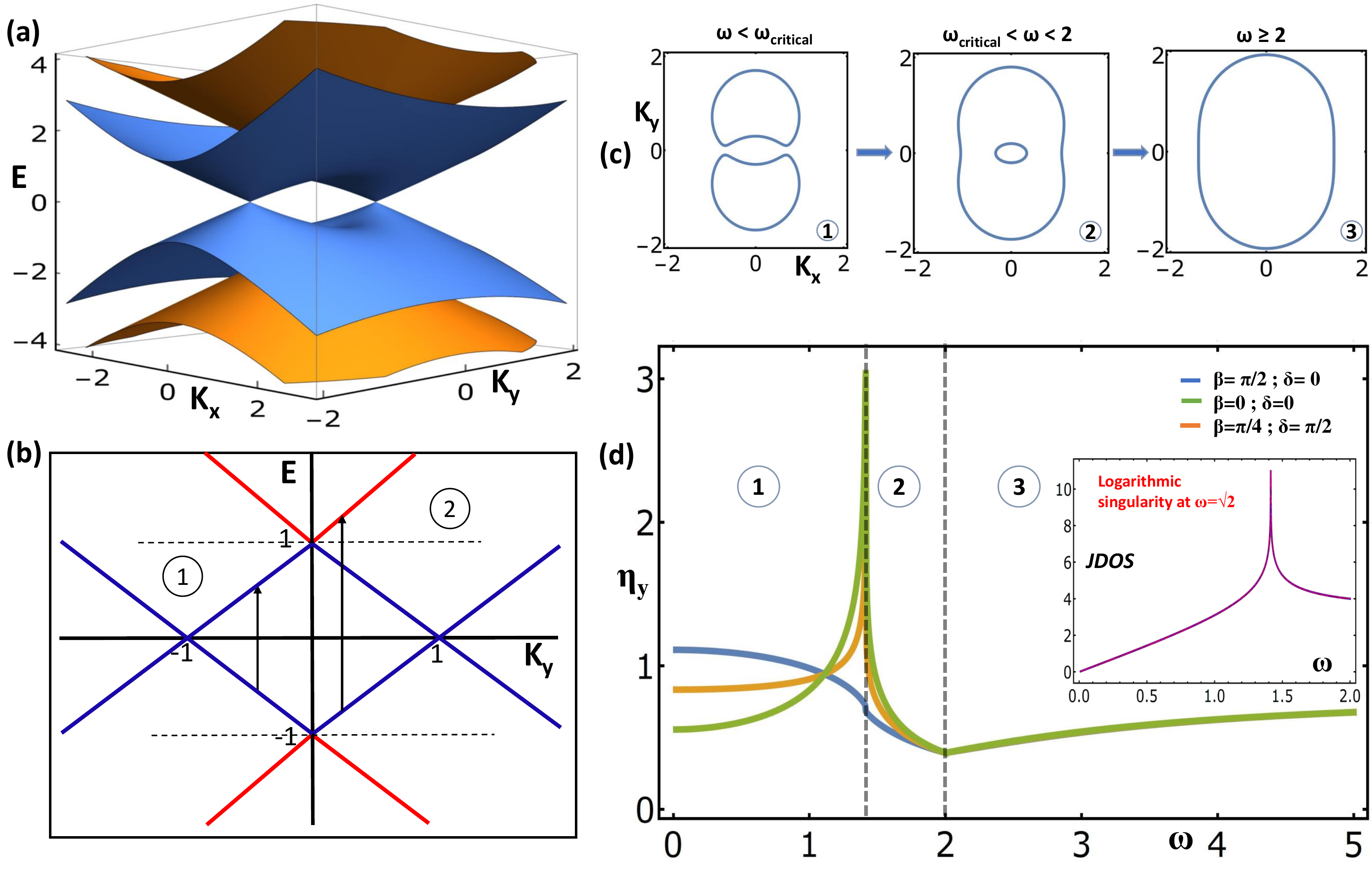}  
    \caption{\small \textbf{(a)} Plot of the energy bands for magnetic field along the $y$-direction for the $\bar X_1$ valley. There is no gap for transitions from $v_1$ to $c_1$, as these two bands touch at the Dirac points $(0,\pm B/v)$. On the other hand, the $v_2$ and $c_2$ bands are separated by a gap $2B$ at $(0,0)$. \textbf{(b)} Plot of the band dispersion along the line $k_x=0$.  Notice that both ``intra-valley" ($v_1\to c_1$)   and ``inter-valley" ($v_1\to c_2$)   transitions are possible, the latter  only for frequencies larger than $2B$. \textbf{(c)} Change of topology of the iso-energy contours with increasing excitation energy. The first change in topology occurs at the saddle-point frequency $\omega=2 B\sin\alpha$, the second at $\omega=2B$. \textbf{(d)} Plot of the absorption coefficient $\eta_y$ (in units of $\eta_{\rm graphene}=\frac{\pi e^2}{\hbar c}$.)  vs. $\omega$ for $\alpha =\pi/4$.  Notice the logarithmic singularity at  $\omega=2 B\sin\alpha$ and the cusp at $\omega=2B$.  These features are related to the changes in the topology of the iso-energy contours shown in panel (c).
      Inset: joint density of states as a function of $\omega$ showing a logarithmic divergence at the saddle point $\omega = 2 B \sin\alpha$, where $\alpha=\pi/4$.} 
    \label{fig:image3}
\end{figure}

Let us start our calculation for the magnetic field in $y$-direction by keeping $\rho=1$ (isotropic valley). This means that the only source of anisotropy is the magnetic field itself. From Fig.~{\ref{fig:image3}}(a) we see that the Dirac point, which  initially was at $\bar {X_1}$, is split by the magnetic field term $B \tau_0 \sigma_y$ into two distinct Dirac points located at $k_x=0$, $k_y=\pm B/v$.  The two Dirac points are related to each other by parity symmetry ($\sigma_0\tau_x$ combined with the inversion of $\kv$), which remains intact in this case.   The bands emanating from the two Dirac cones (depicted in blue in Fig.~{\ref{fig:image3}}(a)) account for the low-frequency part of the absorption spectrum.  The high and low energy bands, depicted in orange in Fig.~{\ref{fig:image3}}(a), become degenerate with the blue bands at the $\bar {X_1}$ point ($k_x=k_y=0$), as required by parity symmetry. Also, this band dispersion gives rise to a saddle point at $k_x= \cos \alpha , k_y=0$ as can be directly verified from Eq.~(\ref{eig_y}). The singularity occurs in the ``blue" bands, and the transition frequency associated with it is $\omega_{\rm critical}=2B \sin\alpha$. 

We performed numerical calculations for the absorption coefficient in the presence of a magnetic field along the $y$-direction. There is no selection rule from the symmetry constraints in this case. Thus we have included all the possible transitions (inter-valley transitions i.e., $v_1 \xrightarrow[]{}c_2$ together with intra-valley transitions i.e., $v_1 \xrightarrow[]{}c_1$) and obtained the optical absorption coefficient as shown in Fig.~\ref{fig:image3}(d). Notice that we only have intra-valley transitions for $\omega \le 2B$. For $\omega \geq 2B$ we include $v_1 \to c_2$ and $v_2 \to c_2$ transitions.

The absorption spectrum has a logarithmic singularity at the saddle-point frequency $\omega = \omega_{\rm critical}=2B \sin\alpha$ (Fig.~\ref{fig:image3}(d)).  Furthermore, this singularity is prominent only for light polarized along the $x$-direction ($\beta=0,\delta=0$).

To understand the origin of this singularity we have calculated the joint density of states (JDOS) which is defined as
\be
g(\omega)= \sum_\kv \delta\big(\omega-E_{c}(\kv)-E_{v}(\kv)\big)= \frac{1}{2\pi\rho v^2}\sum_j \int_0^\infty d\bar k\bar k \int_{\phi_{\rm min}}^{\phi_{\rm max}} d\phi \frac{\delta(\bar k-\bar k_{j}(\phi))}{|d(E_{c}-E_{v})(\bar k,\phi)/d\bar k|}\ ,
\ee
In the vicinity of $\omega_{\rm critical}$ this evaluates to
\begin{equation}\label{JDOS-Singularity}
    g(\omega) \simeq  \frac{\omega}{\pi\rho v^2}\cot\alpha \ln\left\vert\frac{\sin\alpha}{\omega-\omega_{\rm critical}}\right\vert\,.
\end{equation}
Thus we see that the logarithmic singularity in the absorption spectrum for $x$-polarization  reflects an underlying singularity in the JDOS. This singularity can also be understood by observing the  change in the topology of the iso-energy contours which occurs at $\omega_{critical}$, as illustrated in panel (c) of Fig.~\ref{fig:image3}. This effect -- akin to the change of Fermi surface topology in a Lifshitz transition -- is entirely induced by the magnetic field. The downward cusp in the absorption spectrum around $\omega = 2B$ can also be understood  due to the vanishing of the inner iso-energy contour at this excitation frequency, as shown Fig.~\ref{fig:image3}(c(2)).

It remains to be explained why the absorption spectrum does not have a logarithmic divergence at $\omega=\omega_{\rm critical}$ when the incident light is polarized in the $y$-direction. In fact, in this case, we only observe  a small downward cusp at $\omega=\omega_{\rm critical}$.
The reason for this difference becomes clear when we consider the behavior of the squared matrix elements of the current operators ($J_x$ and $J_y$) at the saddle point. The matrix element of $J_y$ vanishes at the saddle point energy, whereas the matrix element of $J_x$ remains finite. This is further discussed in the section IV of the supplementary material\cite{SI}.

The scenario described in the previous paragraphs is not related to band anisotropy, which should be evident from the fact that we observed a strong polarization dependence of the absorption spectrum even though the anisotropy parameter $\rho$  was set to $1$ (we have discussed the optical absorbance for the anisotropic Dirac cone and the total absorbance in the supplementary material section VI \cite{SI}). The scenario is also robust with respect to  variations of the mixing angle $\alpha$: while the plots of Fig.~\ref{fig:image3} have been obtained for $\alpha=\pi/4$, we find the same qualitative behavior for other values of $\alpha$ (section V of the \cite{SI}). 
\section{Summary and discussion}
In this work, we have reported a detailed theoretical analysis of the photon absorption spectrum for a generic model of nonsymmorphic 2D semimetals with and without a magnetic field. In the absence of a magnetic field, the absorption coefficient is very similar to the absorption coefficient for graphene, which is $\frac{\pi e^2}{\hbar c}=2.3\%$ but multiplied by the {\it birefringence} function we define in Eq.~(\ref{biref}). For a strongly anisotropic cone with $\rho \ll 1$, $f(\rho,\beta)$ grows as $1/\rho$, which leads to an order of magnitude enhancement of the absorption coefficient for the anisotropic valley. Furthermore, this enhancement is highly sensitive to the direction of the photon polarization, as can be seen from the $\sin^2\beta$ factor in the {\it birefringence} function. This function together with the photon polarization  can enable a valley selection in absorbance spectrum in the case that a nearly isotropic valley and a strongly anisotropic valley are present near the Fermi level. 

Upon switching on the magnetic field, the absorption spectrum becomes frequency-dependent, and the absorption function is also coupled with anisotropy factor $\rho$ and the mixing angle $\alpha$. We now have more than one control parameter at hand to tune the absorption coefficient, which enables us to tune the absorption coefficient efficiently by varying the photon polarization angle $\beta$. The range of tunability of the total absorbance (taking both isotropic and anisotropic valley into account) is quite broad, going from 1 to up to 10 in units of $\eta_{graphene}$,  which is larger than that envisaged in previous works \cite{mult, Jiang2018}. A magnetic field perpendicular to the plane opens a gap $2B\sin\alpha$, whose observation allows an experimental determination of the hidden "mixing angle" $\alpha$. On the other hand, an in-plane magnetic field leaves the system gapless but splits the Dirac cones and makes absorption more sensitive to photon energy. For example, in Fig.~\ref{fig:image3}(d), we plotted the absorption coefficient for a magnetic field along the $y$-direction as a function of frequency for different photon polarizations. The Van Hove singularity in the joint density of states, from the saddle point in the band dispersion \cite{VHS}, gives rise to a sharp peak in the absorption spectrum and thus greatly enhances the photon absorption at the critical frequency. 

We note that the generic band model of 2D Dirac semimetals is employed in this work, and thus the results can be applied to study the optical properties of all Dirac semimetals with different nonsymmorphic lattice symmetries. The significant tunability in photon absorption enabled by the band anisotropy and external magnetic fields establishes nonsymmorphic 2D Dirac semimetals as a promising platform for optoelectronic and magnetoelectronic device applications.
\begin{figure}[h!]
    \includegraphics[width=0.9\linewidth]{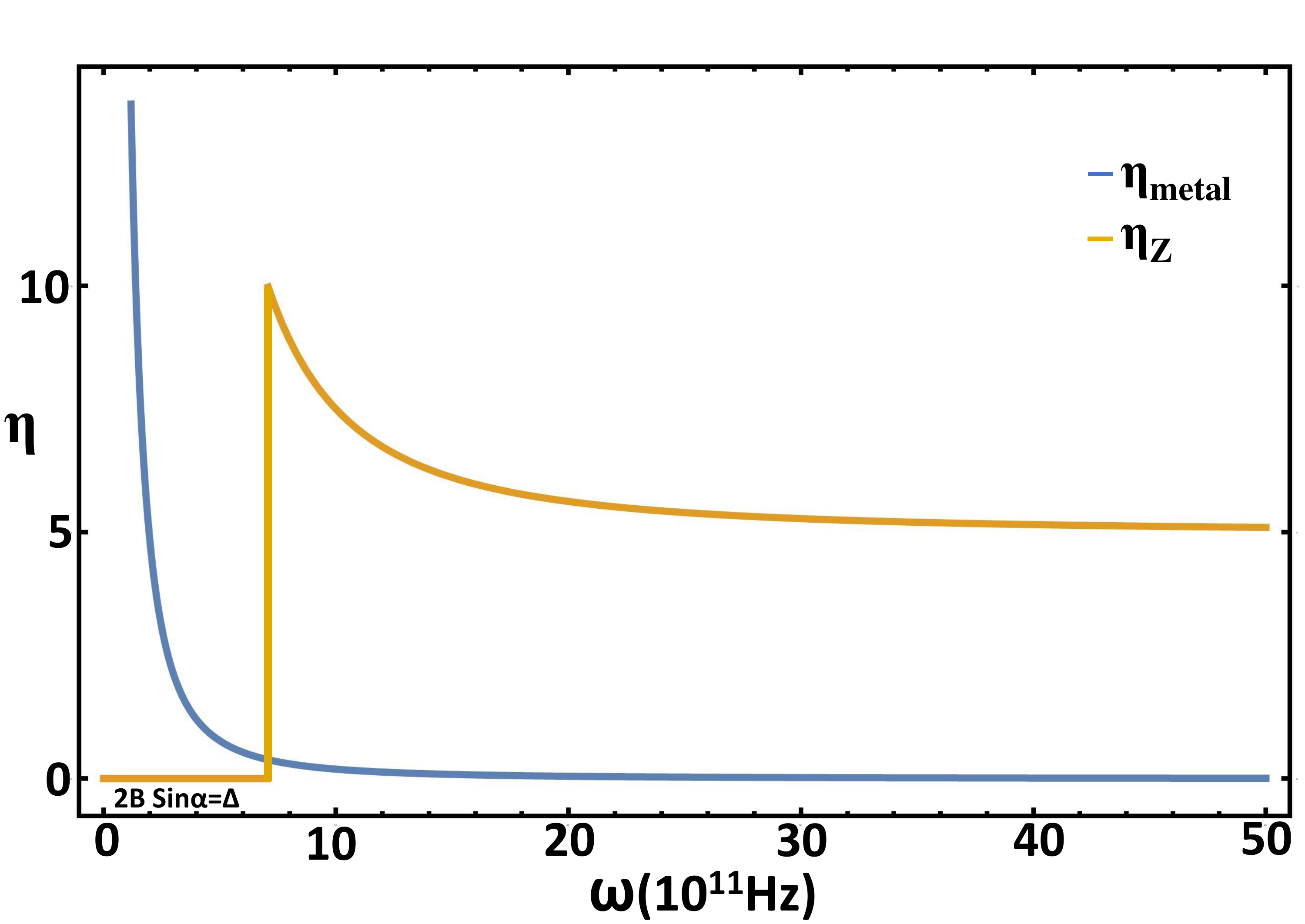}  
    \caption{\small Estimated metallic absorption coefficient ($\eta_{metal}$ -- blue line) vis-a-vis the absorption coefficient due to the Dirac cone with a magnetic field $B$ perpendicular to plane ($\eta_z$ -- orange line) vs. frequency ($\omega$ -- in units of $10^{11}$ Hz.). Our estimate of metallic absorption is based on the Drude formula for metallic conductivity, with Fermi energy on the order of $0.1$ eV and momentum relaxation time $\tau=10^{-10}s$.  We also take B= 5 T and $\alpha=\pi/4$.  For this very clean system the metallic absorption is a Lorentzian with a finite and sharp peak. It is well separated from the Dirac-cone absorption and the latter can be measured with negligible contribution from the former.  The absorption coefficient is in units of $ \frac{\pi e^2}{\hbar c}\simeq 0.023$.} 
    \label{fig:image4}
\end{figure}

We conclude with a few comments on the prospects of observing the features described in this paper in experiments on realistic materials such as $\alpha$-bismuthene. We have already mentioned the problem of positioning the symmetry-protected Dirac cone at the Fermi level, or, at least, very close to it.  This can be achieved by doping or by the application of mechanical stress. However, the proper positioning of the Dirac point is not sufficient to guarantee that the low-frequency portion of the optical absorption spectrum will be dominated by interband transitions between the lower and the upper Dirac cone. The reason is that, in general, there are partially occupied bands giving rise to a low-frequency metallic absorption background, which can obscure the signal from the Dirac cone.  What saves the day is the fact that, in a sufficiently clean system, this background absorption (arising from intraband transitions within the partially occupied bands) has the form of a Drude peak at zero frequency with a width given by the inverse of the momentum relaxation time $\tau$. The bandgap for magnetic field in the $z$-direction is $2B\sin\alpha$, and the photon absorption from interband transition starts above the gap. Assuming a magnetic field of 1 Tesla ($\approx 10^{-23}$Joules) the width of the Drude peak becomes smaller than the gap for $\tau$ greater than $\approx 10^{-11}s$. 

The Drude contribution to the absorption coefficient is calculated by dividing the power dissipated in the layer per unit area, $W_{o} = Re(j.E)=Re(\sigma) E^2$, where $j=\sigma E$ is the current density and $\sigma$ is the Drude conductivity of the carriers in the partially filled band, by the incident power $W_{i}=cE^2/(4\pi)$.
The result is 
\begin{equation}
    \eta_{metallic}= \frac{W_0}{W_i}=\frac{4\pi Re(\sigma)}{c}\,.
\end{equation}
Inserting the well-know expression for the Drude-conductivity of a two-dimensional electron gas,  $\sigma=\frac{\sigma_{0}}{1-i\omega\tau}$, where $\sigma_0=\frac{ne^2\tau}{m}$ and  $n=\frac{m \epsilon_{F}}{\pi\hbar^2}$ is the two-dimensional carrier density and $\epsilon_F$ is the Fermi energy, we arrive at
\begin{equation}
    \eta_{metallic}= \frac{\pi e^2}{\hbar c} \frac{4\epsilon_{F}\tau/\pi \hbar}{1+(\omega\tau)^2}\,.
\end{equation}

This absorption spectrum is plotted in  Fig.\ref{fig:image4}   vis-a-vis the one from the Dirac cone ($\eta_{Z}$). We have assumed a relaxation time $\tau=10^{-10}s$, a Fermi energy of $0.1$ eV, and a magnetic field of 5 T,  which leads to a peak width at half maximum of about ($\mathcal{FWHM}=\frac{2\hbar}{\tau}$) 0.36 Tesla which is significantly smaller than the energy scale ($2B\sin\alpha \approx 7T$) on which the most interesting features of our nonsymmorphic model (for example, the magnetic field-induced van Hove singularity) appear.  Notice that the metallic absorption spectrum is in the form of a Lorentzian ($\frac{A}{\pi} \frac{\tau}{(1+(\omega\tau)^2}$), thus the absorption coefficient remains finite at zero frequency. Choosing a sufficiently large relaxation time we can get the width of the Drude peak to be very narrow. Theoretical studies of two-dimensional materials have shown that large values of the momentum relaxation time are possible \cite{PhysRevB.77.195412, zhan2017distinct}. Therefore we are hopeful that for very clean materials, it will be possible to achieve $\tau\approx {10}^{-10}s$ experimentally. 
Thus, in sufficiently clean materials the spectral features predicted in this paper can be clearly distinguished from the Drude peak arising from the partially occupied bands.

\section*{Acknowledgements}
G.B. was supported by the US National Science Foundation under Grant No. NSF DMR-1809160. 

\bibliography{mybib}
\pagebreak
\setcounter{section}{0}
\setcounter{equation}{0}
\setcounter{figure}{0}
\setcounter{table}{0}
\newcounter{SIfig}
\setcounter{page}{1}
\makeatletter
\renewcommand{\theequation}{S\arabic{equation}}
\renewcommand{\thefigure}{S\arabic{figure}}
\renewcommand{\theSIfig}{S\arabic{SIfig}}
\renewcommand{\bibnumfmt}[1]{[S#1]}
\renewcommand{\citenumfont}[1]{S#1}

\author{Amarnath Chakraborty}
\email{achakraborty@mail.missouri.edu}
\affiliation{Department of Physics and Astronomy, University of Missouri, Columbia, Missouri, USA}
\author{Guang Bian}
\affiliation{Department of Physics and Astronomy, University of Missouri, Columbia, Missouri, USA}
\author{Giovanni Vignale}
\email{vignaleg@missouri.edu}
\affiliation{Department of Physics and Astronomy, University of Missouri, Columbia, Missouri, USA}
\begin{center}
    \textbf{\large Supplemental Materials}\\
\textbf{ Photon Absorption of Two-dimensional Nonsymmorphic Dirac Semimetals}
\end{center}

\maketitle

\section{Construction of the reduced Hamiltonian}\label{appendix:a}

 We can classify the eigenstates according to the eigenvalues of  $\tilde M_z= \tau_y\sigma_z$.  In this section we construct the reduced $2\times 2$ Hamiltonian for the symmetry sectors $\tilde M_z=1$ and $\tilde M_z=-1$.

 In the $\tilde M_z$-even sector (eigenvalue +1) a convenient basis is
 \be
v_1=\frac{1}{\sqrt{2}}\left(\begin{array}{c}
 1\\0\\i\\0
 \end{array} \right)\,,~~~~~v_2=\frac{1}{\sqrt{2}}\left(\begin{array}{c}
 0\\1\\0\\-i
 \end{array} \right)
 \ee
corresponding to $\sigma_z=1,\tau_y=1$ and $\sigma_z=-1,\tau_y=-1$ respectively.
Simple algebra leads to the following reduced Hamiltonian in the  $\tilde M_z$-even sector:
\be
H_{\tilde M_z=1}= \rho vk_x (\cos \alpha \sigma_z+\sin \alpha \sigma_x) + v k_y \sigma_y
\ee
The eigenvector of  $H_{\tilde M_z=1}$ are best expressed in the basis 
\be\label{basis}
\frac{v_1+iv_2}{\sqrt{2}}=\frac{1}{2}\left(\begin{array}{c}
 1\\i\\i\\1
 \end{array} \right)\,,~~~~~\frac{v_1-iv_2}{\sqrt{2}}=\frac{1}{2}\left(\begin{array}{c}
 1\\-i\\i\\-1
 \end{array} \right)
 \ee
In this basis (i.e., interpreting $\frac{v_1+iv_2}{\sqrt{2}}$ as the conventional $|\up\rangle$ and $\frac{v_1-iv_2}{\sqrt{2}}$ as the conventional $|\down\rangle$),  $\sigma_z$ becomes $\sigma_x$, $\sigma_x$ becomes $\sigma_y$, and $\sigma_y$ becomes $\sigma_z$. Thus we have
\be
 \tilde H_{\tilde M_z=+1}= \rho v k_x (\sigma_x \cos\alpha  \ +  \sigma_y \sin\alpha) + v k_y\sigma_z
\ee

Similar manipulations for the $\tilde M_z$-odd sector (eigenvalue -1) lead to
\be
 \tilde H_{\tilde M_z=-1}= \rho v k_x (-\sigma_x \cos\alpha  \ +  \sigma_y \sin\alpha) + v k_y\sigma_z \,.
\ee

\section{Calculation of the band gap}
To find the gap, let us first minimize the Eq.~(\ref{Ephi}) with respect to $\bar k$.  The minimum occurs at $\bar k =  \mp B F(\phi)$ and substituting this in the above expression we find the angle-dependent energy
\be
\bar E(\phi)=\pm B\sqrt{1-F^2(\phi)}\,,
\ee
where
\be
F^2(\phi) = (\cos\phi\cos\alpha)^2+\sin^2\mu (\cos\phi \cos \nu \sin\alpha+\sin \phi \sin\nu)^2\,.
\ee
The last step is to minimize $\bar E(\phi)$, which means maximizing $F^2(\phi)$ with respect to $\phi$.
First notice that
\be
(\cos\phi \cos \nu \sin\alpha+\sin \phi \sin\nu)^2 =  (\cos^2\nu\sin^2\alpha+\sin^2\nu)\cos^2(\phi+\gamma)
\ee
where
\be\label{TanGamma}
\tan \gamma= -\frac{\tan \nu}{\sin\alpha}\,.
\ee
Making use of this we rewrite
\ber\label{F2}
F^2(\phi) &=& A\cos^2\phi+C \cos^2(\phi+\gamma)\nn\\
&=&\frac{1}{2}\left\{A+C+A\cos(2\phi)+C\cos[2(\phi+\gamma)]\right\}\,.
\eer
where
\be
A=\cos^2\alpha\,,~~~~~C= \sin^2\mu(\cos^2\nu\sin^2\alpha+\sin^2\nu)
\ee
Setting the derivative of $F^2(\phi)$ with respect to $\phi$ to zero we get
\be
\frac{dF^2(\phi)}{d\phi}=-A\sin(2\phi)-C\sin[2(\phi+\gamma)]=0\,.
\ee
which gives us
\be
\tan(2\phi)=-\frac{C\sin(2\gamma)}{A+C\cos(2\gamma)}\,,~~~~~\tan[2(\phi+2\gamma)]=\frac{A\sin(2\gamma)}{C+A\cos(2\gamma)}
\ee
From this we get
\ber
\cos(2\phi)&=&\frac{A+C\cos(2\gamma)}{\sqrt{A^2+C^2+2AC\cos(2\gamma)}}\nn\\
\cos[2(\phi+\gamma)]&=&\frac{C+A\cos(2\gamma)}{\sqrt{A^2+C^2+2AC\cos(2\gamma)}}
\eer
Finally, substituting this in Eqs.~(\ref{F2}) and Eq.~(\ref{Ephi}) we get the gap $\Delta$ given by
\be
\frac{\Delta}{B} = 2\min \frac{|\bar E(\phi)|}{B}=2\sqrt{1-\frac{A+C+\sqrt{A^2+C^2+2AC\cos(2\gamma)}}{2}}\,,
\ee
which is the same expression we got in Eq.~(\ref{Gap1}).
\section{Calculation of the constant energy contours}
With $\bar k=\sqrt{\bar k_x^2+\bar k_y^2}$, $\bar k_x=\rho v k_x/B = \bar k\cos\phi$, $\bar k_y=v k_y/B=\bar k \sin\phi$, and setting $B=1$ as the unit of energy we define the conduction band energies as 
\be
E_{c1}(\bar k,\phi)=\sqrt{1+\bar k^2-2\bar k F(\phi)}\,,~~~E_{c2}(\bar k,\phi)=\sqrt{1+\bar k^2+2\bar k F(\phi)}
\ee
and $E_{v1}=-E_{c1}$, $E_{v2}=-E_{c2}$.  The $\alpha,\mu,\nu$-dependence of $F$ is omitted. Notice that $\bar k>0$ and $0<F<1$ so we have  $E_{c1}<E_{c2}$ for all $\kv$.

The constant excitation energy contour for $v1 \to c1$ transitions is
\ber
(v1 \to c1): \bar k_+(\phi) &=& F(\phi)+\sqrt{F(\phi)^2 + \left (\frac{\omega}{2}\right)^2-1}\nn\\
\bar k_-(\phi) &=& F(\phi)-\sqrt{F(\phi)^2 + \left (\frac{\omega}{2}\right)^2-1}\nn\\
&&\Delta<\omega<2\,,~~~\phi_{\rm min} <|\phi|<\phi_{\rm max}\,.
\eer
where $\omega$ is the excitation energy and the angles $\phi_{\rm min}$, $\phi_{\rm max}$ are the positive solutions of the equation
\be
F(\phi)^2=1-\left (\frac{\omega}{2}\right)^2\,.
\ee 
The constant energy contour for $v1\to c2$ (or $v2\to c1$) transitions is
\ber
(v1 \to c2): \bar k(\phi) &=& \frac{\omega}{2}\sqrt{\frac{\omega^2-4}{\omega^2-4 F(\phi)^2}}\nn\\
&&\omega>2
\eer
This is simply a closed loop around the origin.

Finally, the constant excitation energy contour for the $v2 \to c2$ transitions is
\ber
(v2 \to c2): \bar k(\phi) &=&-\bar k_-(\phi)= -F(\phi)+\sqrt{F(\phi)^2 + \left (\frac{\omega}{2}\right)^2-1}\nn\\
&&\omega>2\,.
\eer
Again, this is a closed loop around the origin.
\section{Behavior of ($|M|^2_{x}$) and ($|M|^2_{y}$) at the saddle point}
From Fig.~\ref{fig:spresults1} we can see that the squared matrix element of $J_y$ ($|M|^2_{y}$) goes exactly to zero at the saddle point ($k_x=1/\sqrt{2}$ and $k_y=0$ for $\alpha=\pi/4$), thus suppressing the effect of the singularity in the joint density of states. At the same time, the squared matrix element of $J_x$ (($|M|^2_{x}$)) remains finite, and this is why the singularity of the joint density of states is fully reflected in the absorption spectrum (Fig.~\ref{fig:image3}(d)) for polarization along $x$.
\begin{figure}[ht]
  \centering
    \includegraphics[width=0.70\linewidth]{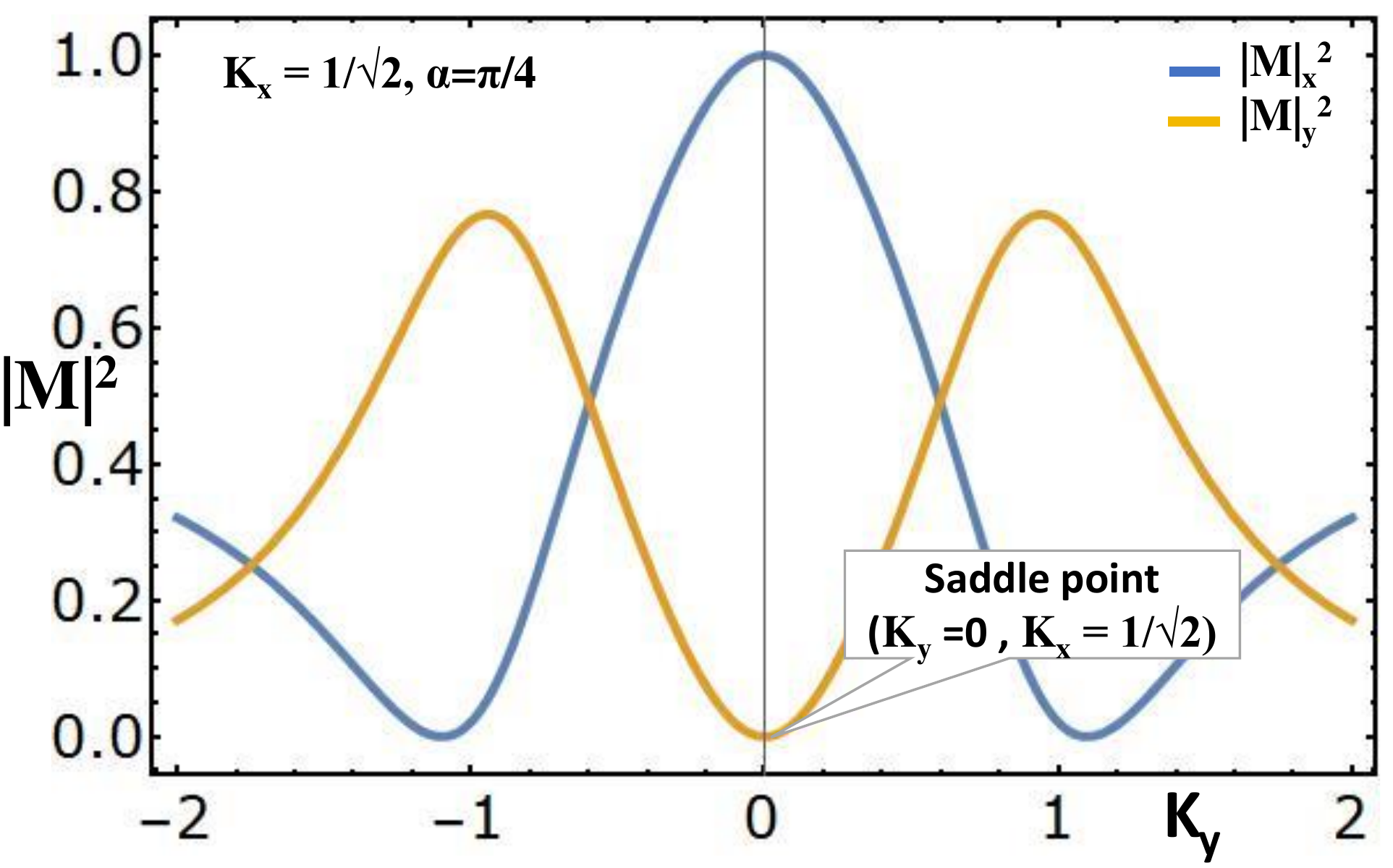}  
    \caption{\small Plots of the squared matrix elements of $J_x$ and $J_y$ (blue and orange lines, respectively) vs $k_y$ at constant  $k_x =\cos\alpha= 1/\sqrt{2}$ for $\alpha=\pi/4$.  Observe that $|M|^2_{y}$ (orange) goes to zero at the saddle point ($k_x=1/\sqrt{2},k_y=0$), while  ($|M|^2_{x}$) (blue) tends to $1$.}
\end{figure}\refstepcounter{SIfig}\label{fig:spresults1}
\section{Dependence of the absorption spectrum on mixing angle}\label{var_alpha}
For magnetic field along the $y$-direction we have checked that the behavior of the  absorption coefficient for $\alpha=\pi/3$ and $\alpha=\pi/5$ is qualitatively similar to what we found for $\alpha=\pi/4$ in the main text. 
\begin{figure}[ht]
  \centering
    \includegraphics[width=1.0\linewidth]{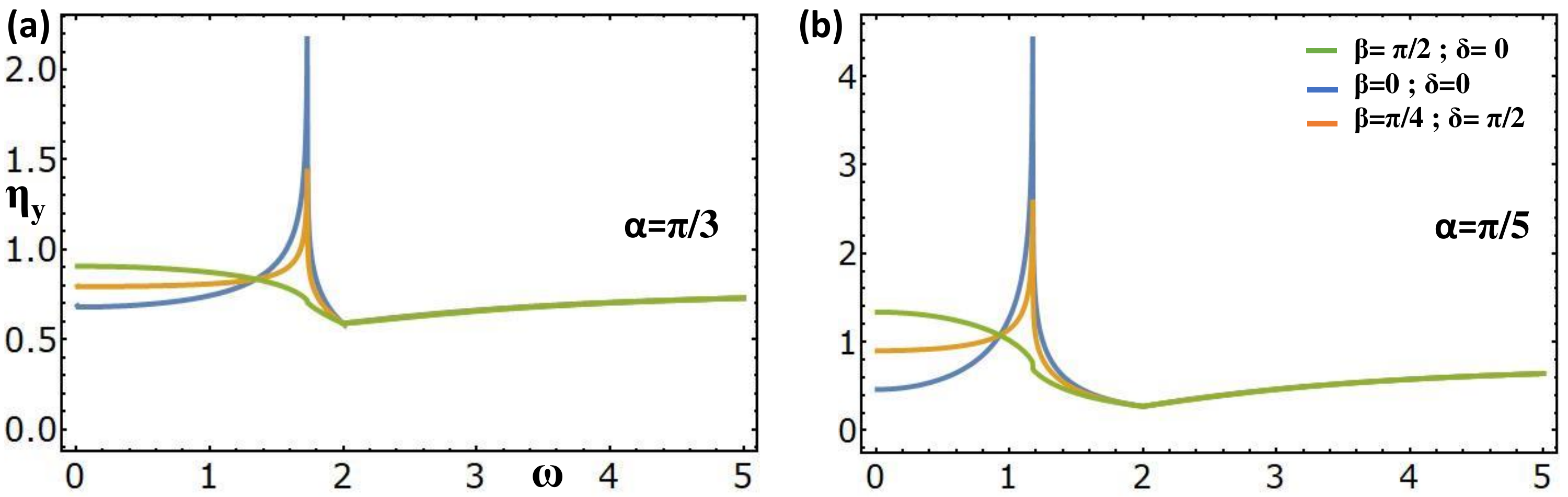}   
    \caption{\small Optical absorption coefficient vs frequency  for magnetic field in the $y$-direction for two values of $\alpha$ different from the one considered in the text \textbf{(a)} $\alpha=\pi/5$ and \textbf{(b)} $\alpha=\pi/3$. Notice that the singular behavior of the absorption coefficient at the saddle point singularity is present at all $\alpha$ with the same polarization dependence that was discussed in the main text. $\rho=1.0$.}
\end{figure}\refstepcounter{SIfig}\label{fig:spresults2}

Fig.~(\ref{fig:spresults2}) shows that, with increasing $\alpha$, the peak in the absorption spectrum shifts to higher frequency, tracking the position of the van Hove singularity  $\omega_{critical}= 2B \sin\alpha$, but the qualitative behavior of the  spectrum near the singularity, including its dependence on polarization, remains unchanged.
\section{Dependence of the saddle-point singularity on the
anisotropy parameter \textbf{$\rho$}}
In $y$-direction of magnetic field if we want to consider both the isotropic and anisotropic valley then the absorption coefficient takes the form 
\begin{equation}
    \eta_y = \Big( \rho\ \cos^2\beta \ \eta_{y}(x) + \rho^{-1} \sin^2\beta \ \eta_{y}(y) \Big)
\end{equation} 
where $\eta_{y}(x)$ and $\eta_{y}(y)$ are the values of $\eta_y$ for $x$ and $y$ directions of polarization of the incident light.

To better understand the interplay between the saddle-point singularity and the anisotropy parameter, we have calculated the optical absorption coefficient for magnetic field in the $y$-direction for $\rho=0.1$, which means the Dirac cones are now strongly anisotropic. 

\begin{figure}[ht]
  \centering
    \includegraphics[width=1.0\linewidth]{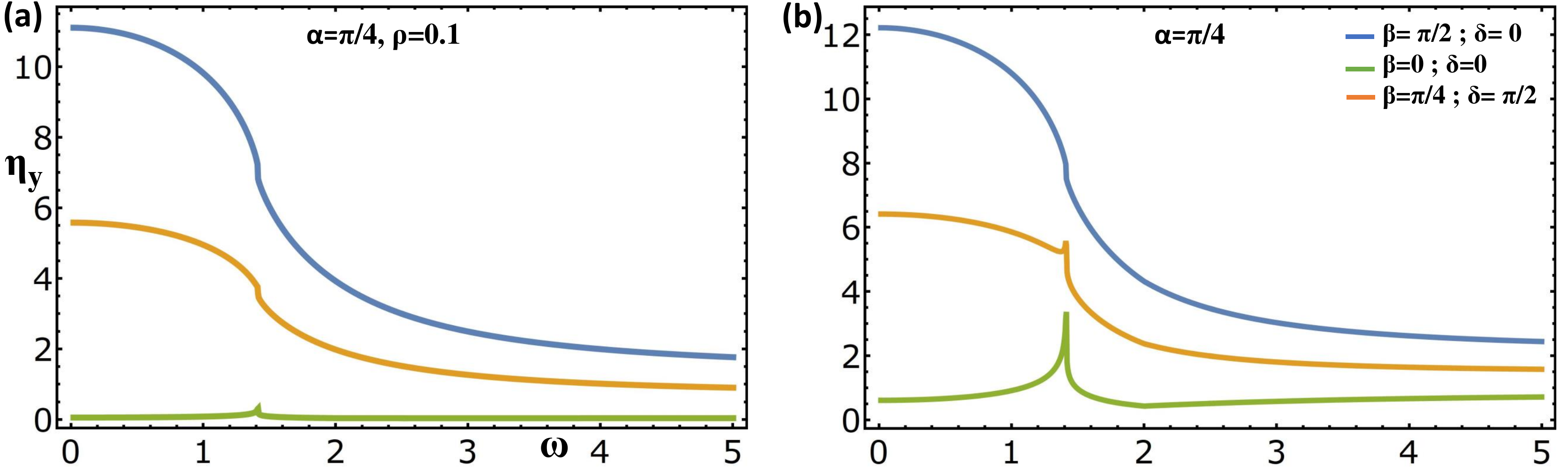}   
    \caption{\small \textbf{(a)} Plot of  the optical absorption coefficient vs $\omega$ for the anisotropic valley ($\rho=0.1$). \textbf{(b)} This figure shows the total optical absorbance. By changing the polarization of the incoming light, we can tune the total absorption coefficient. Both figures are in units of $\eta_{\rm graphene}$ for  magnetic field in the $y$-direction, and $\alpha=\pi/4$.}
\end{figure}\refstepcounter{SIfig}\label{fig:spresults3}
The results are plotted in Fig.~\ref{fig:spresults3}(a) for three different polarizations of the incident light. From this figure we can see that, while the saddle point singularity is still present when the incident light is polarized in the $x$ direction (green curve), it becomes quantitatively small, and it is hardly visible on the scale of the graph. 

This shows that one can effectively play the band anisotropy and the saddle-point singularity against each other, thus adding another dimension to the polarization dependence of the absorption spectrum. The figure in panel (b) of Fig~\ref{fig:spresults3} shows the total optical absorption coefficient $\eta_y(\bar X_i)+\eta_y(\bar X_a)$ for magnetic field along the $y$-direction. Notice that the absorbance can change by an order of magnitude upon rotating the polarization angle.
\section{Nonsymmorphic symmetry in a BISMUTH MONOLAYER STRUCTURE}
The figure \ref{fig:spresults4}shows the crystal structure and band structure of nonsymmorphic Bi-monolayer. \cite{bismuth,PhysRevB.96.205434}. 
\begin{figure}
  \centering
    \includegraphics[width=1.0\linewidth]{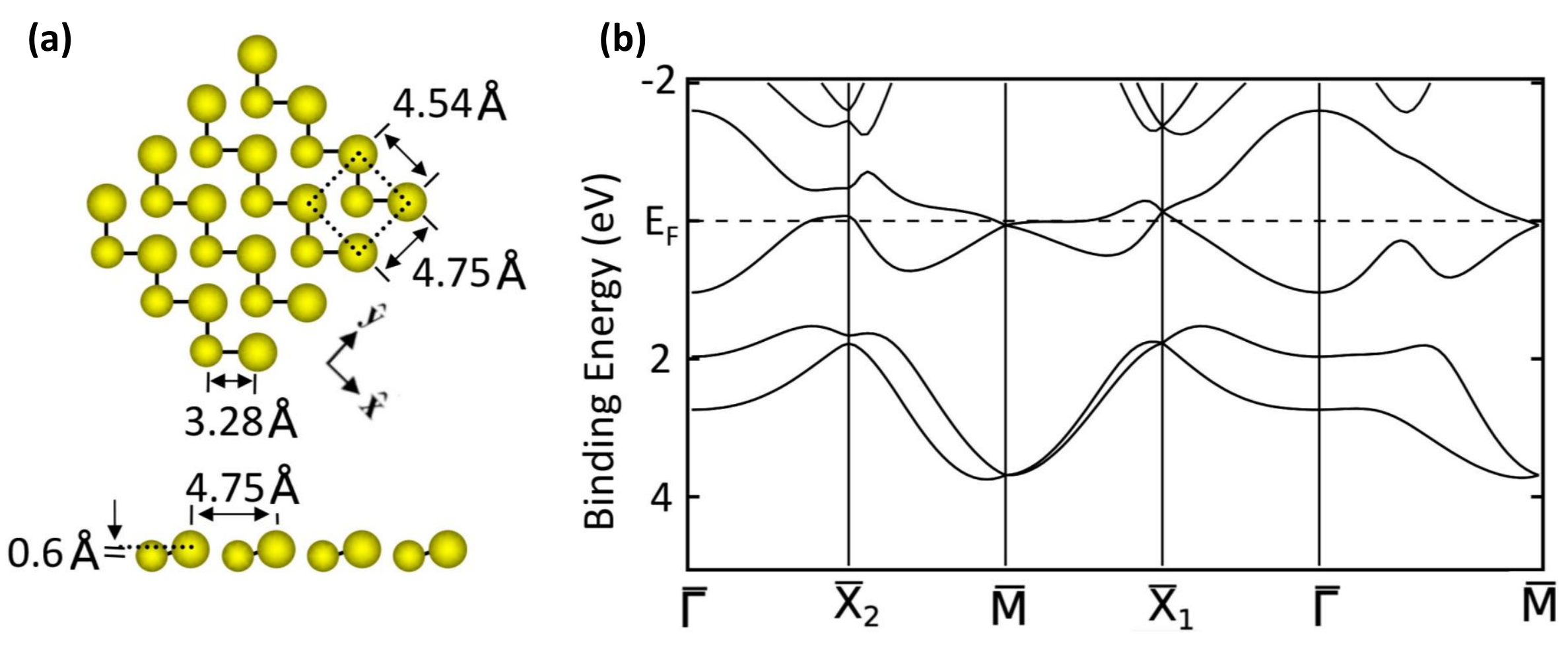}   
    \caption{\small \textbf{(a)} The top and
side views of Bi monolayer lattice structure. \textbf{(b)} DFT band structure of Bi monolayer. the approximate value of the Fermi energy here is 0.1eV.}
\end{figure}\refstepcounter{SIfig}\label{fig:spresults4} 
We can see the two atoms in the unit cell are not exactly in  the same plane from this figure. The layer group has two generators (with spin matrices) as follows:
\begin{equation}\label{generators2}
        \begin{aligned}
        \tilde C_{2x}: (x-1/2,-y,-z)\ i\sigma_x\ , \\  
        P: (-x,-y,-z)\ \sigma_0\ .
    \end{aligned}
\end{equation}
where $\tilde C_{2x}$ is a nonsymmorphic 2-fold screw axis operation. The nonsymmorphic symmetry-protected Dirac states appear very close to the Fermi level at $\bar M$ and $\bar X_1$.

\end{document}